\documentclass[journal]{IEEEtran}
\usepackage{multirow}
\usepackage{changepage}
\usepackage{color}

\usepackage{mathrsfs}
\usepackage{pifont}
\usepackage{bbding}
\usepackage{tipa}
\usepackage{cite}
\usepackage{epsfig}
\usepackage{graphicx}
\usepackage{url}
\usepackage{amsfonts}
\usepackage{multicol}
\usepackage{float}
\usepackage{times}
\usepackage{psfrag}
\usepackage{subfigure}
\usepackage{stfloats}
\usepackage{footnote}
\usepackage{array}
\usepackage{amsmath,epsfig}
\usepackage{stmaryrd}
\usepackage{amssymb}
\usepackage{epic}
\usepackage{curves}
\usepackage{bm}
\usepackage{balance}
\usepackage{caption}
\usepackage{subfloat}
\usepackage{subfig}
\usepackage{graphicx}
\begin{document}

\title{Microwave Surveillance based on Ghost Imaging and Distributed Antennas}
%\author{Xiaopeng~Wang, Zihuai~Lin}

\author{Xiaopeng~Wang,~\IEEEmembership{Student~Member,~IEEE,}and
        Zihuai~Lin,~\IEEEmembership{Senior~Member,~IEEE}
\thanks{The authors are with the School of Electrical and
Information Engineering, The University of Sydney, NSW, 2006,
Australia (e-mail:
\{xiaopeng.wang,zihuai.lin\}@sydney.edu.au).}
}

\maketitle

\begin{abstract}

In this letter, we proposed a ghost imaging (GI) and distributed antennas based microwave surveillance scheme. By analyzing its imaging resolution and sampling requirement, the potential of employing microwave GI to achieve high-quality surveillance performance with low system complexity has been demonstrated. The theoretical analysis and effectiveness of the proposed microwave surveillance method are also validated via simulations.

%In order to overcome the low performance, short range and high complexity that exist in conventional monitoring systems, ghost imaging, originated from the optical area, is adopted into microwave surveillance applications using distributed antennas. Principles about the sampling requirement and imaging limitation of the proposed scheme are discussed in this letter, demonstrating its potential in achieving high-resolution large-scale surveillance performance with low system complexity. Furthermore, the verification of the corresponding theoretical discussions and the effectiveness of the proposed massive surveillance scheme are also presented via simulation results.
\end{abstract}

\begin{IEEEkeywords}
Microwave surveillance, microwave ghost imaging, distributed antennas, chaotic modulated signals.
\end{IEEEkeywords}

%=======================================================introduction=======================================================
\section{Introduction}\label{introduction}
%Massive surveillance, especially in urban environments, is an essential technology with numerous applications in police missions, disaster rescues and military operations. Existing surveillance systems could offer continuous monitoring, but they highly rely on cameras\cite{6869527} or UAVs\cite{6870939} that are easy to be evaded and effected by changing in the climate and illumination. Moreover, optical surveillance methods are sensitive to the changing in climate and illumination, making it not suitable for all-weather monitoring tasks. Moreover, there are other approaches using microwave signals for urban sensing\cite{6898877} and hidden target detection\cite{6906269}, their short range, bulky size and limited performance are still not ideal for practical cases that requires large-scale, low-complexity and high-resolution surveillance capability to provide constant situation awareness.

Surveillance is an essential technology with numerous applications in police missions, disaster rescues and military operations. Existing optical surveillance systems are commonly used in practise, but they highly rely on cameras\cite{6870939} or UAVs\cite{6869527}. Therefore, their performance will be easily affected by climate and illumination conditions. Due to the reason that microwave signals possess the ability of penetration and are insensitive to environmental factors, the possibility of employing microwave signals for surveillance has drawn much attention and been widely discussed in the literature \cite{viani2011localization}. For instance, based on RSSI measurements in a WSN area, the authors in \cite {viani2008object} and \cite{viani2010electromagnetic} presented a classification approach for passive target localization and tracking. By adopting radar imaging techniques, hidden target detection and imaging methods are proposed in \cite{6898877} and \cite{6906269} for urban scenarios as well.

Ghost imaging (GI) is a high-resolution and nonlocal imaging method derived from the optical area. In 1995 \cite{pittman1995optical}, the first experiment of GI was implemented under laboratory environment with entangled photons, then successfully completed by using classic thermal light sources in 2004 \cite{gatti2004ghost} as well. The basic mechanism of GI is to reconstruct the target image by calculating the second order correlation between the reference optical field and the signal received by a single pixel detector. Later, by applying compressive sensing (CS) theory into GI, the algorithm proposed in \cite{jiying2010high} not only reduced at least 10 times of the measurements to achieve high resolution reconstruction results, but also contributed a brand new imaging algorithm besides the traditional one. The latest GI algorithm proposed in 2014 \cite{zhang2014object} reconstructs original objects by computing the the pseudo-inverse of the matrix constituted by the row vectors of each speckle field, providing a further improvement in reducing the number of detections, computing time and increasing the imaging quality.

In this letter, in order to achieve high-quality performance with low system complexity, we introduce GI mechanism into microwave surveillance scenarios. Based on our preliminary work\cite{ICASSP2015}, GI's characteristics in imaging resolution and sampling requirement are further analyzed, then a GI and distributed antennas based microwave surveillance method are proposed. Meanwhile, chaotic modulated signals are employed to generate spatially incoherent electromagnetic (EM) fields for illumination, which are later processed with uniquely sampled reflected signals captured by a single receiving antenna for image reconstruction. In addition, simulation results are also presented to verify the theoretical analysis and evaluate the effectiveness of the proposed surveillance method under different SNR levels.

%In this letter, in order to satisfy the requirement of large-scale surveillance, we introduce the GI mechanism into microwave detection scenarios. Theoretical characteristics of GI are analyzed and proven to be suitable for high-resolution microwave surveillance using distributed antennas. Meanwhile, chaotic modulated signals are used to generate spatial incoherent electromagnetic (EM) fields at the target plane, which are later processed with uniquely sampled reflected signal captured by a single receiving antenna for image reconstruction. Simulation results are also presented to verify the theoretical discussions and effectiveness of the proposed massive surveillance scheme.

%This novel massive surveillance scheme possesses remarkable significance. The sub-antennas in the system do not need to be restricted to any specific types. Instead, antennas distributed in existing systems such as cellular, Wi-Fi and radar can all be employed to perform the proposed scheme. This feature not only ensures the large coverage, but also the reliability. It is difficult to be detected, evaded or destroyed. It is ideal for conceal monitoring, and tactical reconnaissance in urban scenarios. Besides, the proposed scheme can be used in disaster rescues as well. Robots that are randomly sowed over the debris can be treated as sub-antennas and organized to perform large-scale searching for survivors.

The organization of the remainder of this letter is as follows. In Section \ref{theoreticalAna}, theoretical principles are presented about the imaging resolution and sampling requirement of microwave GI. In Section \ref{systemModel}, the typical scenario of the proposed GI and distributed antennas based microwave surveillance is mathematically presented. In Section \ref{algorithm}, the imaging algorithm of is proposed. Then in Section \ref{simulation}, simulation results are shown to validate the theoretical analysis in Section \ref{theoreticalAna} and to demonstrate the effectiveness of the proposed scheme. Finally, in Section \ref{conclusion}, some concluding remarks are drawn. 
%=====================================================================================================================================================
\section{Motivation on Applying microwave GI }\label{theoreticalAna}
Performing GI using microwave signals has been demonstrated to be suitable for long-range detections \cite{6518188}\cite{7031421} and through-wall imaging applications\cite{ICASSP2015}. In this section, based on these previous research, we further deduce the feasibility of applying GI into microwave surveillance scheme using distributed antennas. Since the quality of reconstruction results of GI relies on the independence in time and space of detecting fields and received signals\cite{6518188}, the coherence size and sampling interval are two major aspects to be discussed here.
%===================================================================
\subsection{Coherence Size}
In order to achieve high-resolution GI reconstruction results, the details of the targets should be larger than the size of the coherence area of detecting fields. The reflected signals containing details of the object within the range of coherence area will be significantly correlated to each other, causing difficulties in achieving high-resolution GI reconstruction result \cite{gatti2006coherent}.

In optical cases where there are two point sources at the two ends of a linear source, assuming the distance between the two point sources is $b$, the distance between the center of this linear source and the target plane is $R$, then for 2-D scenarios, the coherence size $d_c$ on detecting fields generated by the linear optical source can be expressed as \cite{goodman2007speckle},

\begin{equation}
d_c=\frac{R\lambda}{2b}.
\end{equation}

While for practical optical linear sources, where there are a large amount of other point sources between the above two, the coherence size $d_c$ on detecting fields can be expressed as \cite{goodman2007speckle},

\begin{equation}
d_c=\frac{R\lambda}{b}.
\end{equation}

As a result, in our case which can be represented by a similar linear source model containing sparse microwave radiation elements as point sources, the range of coherence size $d_c$ on the corresponding detecting fields can be expressed as,

\begin{equation}\label{rangeCoherence}
\frac{R\lambda}{2b}\leq d_c \leq \frac{R\lambda}{b}.
\end{equation}

Therefore, the coherence size, or the the imaging resolution of GI is determined by three factors, namely the carrier frequency, the size of the microwave source and the distance from the source to the imaging area. Obviously, in our scenario where the wavelength $\lambda$ is much larger than optical signals, and the distance $R$ is normally fixed, it is natural to employ distributed antennas to form an incoherent fluctuation source with a large size to provide refined reconstruction performance.

\subsection{Sampling Interval}
%In traditional signal processing, Nyquist's sampling theorem should be obeyed. For any signal, the sampling rate should be at least two times lager than the bandwidth of the signal to avoid distortions. However, in the GI mechanism, the sampling interval should satisfy the condition that is much larger than the signal coherence time. Any sampling that falls within the coherence time will not be fully independent to the nearby ones along the time sequence. This will introduce unexpected coherence and reduce the independence in time, and therefore lead to the failure in obtaining high-resolution GI results\cite{ferri2005high}.

\begin{table}[!t]
\renewcommand{\arraystretch}{1.3}
\caption{Sampling Requirement Comparison}
\label{comparisonSampling}
\centering
\begin{tabular}{|c|c|c|}
\hline
\ &\bfseries Nyquist's Sampling Theorem & \bfseries GI Sampling Requirement\\
\hline
\ \bfseries Purpose & Avoid Aliasing & Avoid Coherence\\
\hline
\ \bfseries Sampling Rate & $\gg$ Signal Bandwidth & $\ll$ Signal Bandwidth \\
\hline
\end{tabular}
\end{table}

In the previous research of microwave GI implementations\cite{6518188}\cite{7031421}, the sampling rates are set to match Nyquist's sampling theorem. However, this setting of parameters may need to be adjusted. In the GI mechanism, samplings that with too narrow intervals will not be fully independent to the nearby ones along the time sequence. For the purpose of avoiding unexpected coherence and ensuring the independence in time, the sampling interval should satisfy the condition that is much larger than the signal coherence time\cite{ferri2005high}.

%fall the sampling interval should satisfy the condition that is much larger than the signal coherence time. Any sampling that falls within the coherence time will not be fully independent to the nearby ones along the time sequence. This will introduce unexpected coherence and reduce the independence in time, and therefore lead to the failure in obtaining high-resolution GI results\cite{ferri2005high}.

Assuming the bandwidth of the detecting signal is $B$, then its coherence time $\Delta t$ can be expressed as\cite{goodman2007speckle},
\begin{equation}\label{coherenceEqua}
\Delta t=\frac{1}{B}.
\end{equation}

Apparently, the larger the signal bandwidth is, the shorter its coherence time is, and the lower the corresponding sampling rate should be applied.

This sampling requirement is different from, but does not contradict to the well-known Nyquist's sampling theorem. It is because the purpose of sampling in GI is to get the certain signal strength at certain time points along the time sequence, not to fully recover the whole period of signal without distortion. Therefore, this feature remarkably reduces the system complexity and cost, providing the precondition of the widely implementation of the proposed distributed antennas and GI based microwave surveillance scheme as well.

A comparison between Nyquist's sampling theorem and requirement of sampling for GI is shown in Table \ref{comparisonSampling}.

%However, it should be mentioned that. That participants in the procedure of GI reconstruction is not the signal amplitude with positive or negative signs. Instead, it is the signal strength, or the intensity received by optical sensors or antennas that is used to retrieve the image of objects. As a result,

%This feature provides the precondition of the widely implementation of microwave GI and distributed antennas based surveillance scheme, since the requirement of system complexity and cost has been reduced remarkably.

%===================================================================================================================
\section{System Model}\label{systemModel}
\begin{figure}
\centering
\includegraphics[width=2.5in,angle=0]{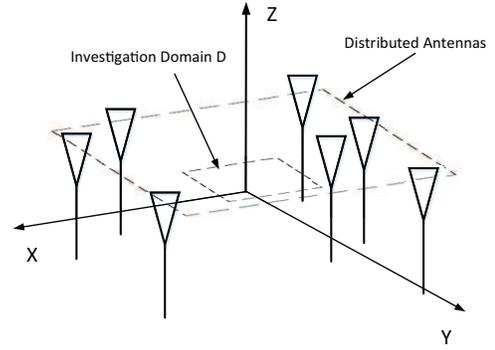}
\caption{A typical scenario of microwave surveillance based GI and distributed antennas.}
\label{scenario}
\end{figure}

A typical scenario of the proposed surveillance scheme using microwave GI and distributed antennas is shown in Fig.\ref{scenario}. The investigation domain D is located in the XY-plane. It is inhomogeneous, containing free space and several obstacles to be imaged. The targets in D are identified by the 2-dimensional (2D) distribution of scattering coefficients $\delta_{r_{o}}$,$r_{o}\in D$, where $r_o$ is the position vector. The scenario under measurement is illuminated by a set of distributed antennas, which are deployed with the same distance from D along the Z-axis, but at different spatial locations $r_{i}$,$i=1,2,\ldots, I$ The $ith$ transmitting antenna radiates an electromagnetic signal $S_i$. In addition, the receiving antenna is fixed at the location $r_r$, with the same distance from D.

%Several sub-antennas are deployed distributively, transmitting statistically independent signals, and forming an equivalent microwave source for detection. The region to be imaged is set to be located parallel to the plane of distributed antennas. The reflected signal from objects is collected by a single receiving antenna, then processed by using GI reconstruction algorithm to retrieve the image of the target plane.

%It consists of several sub-antennas forming as a microwave source, the target imaging plane and the free space. The region to be imaged is located parallel to the plane of the distributed antennas. The reflected signal is collected by a single receiving antenna, then processed by using GI reconstruction algorithm to retrieve the image of the target plane.

\section{microwave GI reconstruction algorithm}\label{algorithm}
Consider a single point target located in the free space under the illumination from the distributed antennas, the reflected signal can be expressed as,

\begin{equation}\label{reflected}
r(t)=\delta \sum_{i=1}^I S_{i}(t- \tau_{i}),
\end{equation}where $\delta$ is the scattering coefficient of the single point target, and $\tau_{i}$ is the propagation delay \cite{6747822}.

Chaos waveforms share the same statistically-independent feature with random noise, but they are easier to be generated and controlled\cite{ICASSP2015}. Therefore, in our case the $S_{i}(t)$ in equation (\ref{reflected}) represents the chaotic modulated signal transmitted by the $ith$ transmitting antenna, which can be generated using the following cubic mapping sequence,

\begin{equation} \label{chaos}
x_{i}(n+1)=a(x_{i}(n))^3-(1-a)(x_{i}(n)),
\end{equation}where $x_{i}(0)$ is the initial value, and the coefficient $a$ is normally restricted to the range $0<a\leq4$ \cite{rogers1983chaos}. Then, the chaotic amplitude modulated signal can be written as,

\begin{equation}
S_{i}(t)=Amp_{i}(t) \exp(2\pi ft+\varphi),
\end{equation}where $f$ is the frequency and $\varphi$ is the phase,

\begin{equation}\label{amp}
Amp_{i}(t)=\sum_{n}x_{i}(n) g(t-n\Delta t).
\end{equation}

In equation (\ref{amp}), $g(t)$ is the Dirac Delta function, $\Delta t$ is the discrete sampling interval and $x_{i}(n)$ is the nth value of the cubic sequence described by equation (\ref{chaos}).

Extend the above single target model and divide the whole investigation domain into a finite number of sub-planes with P rows and Q columns. Each sub-plane can be treated as a single point target, whose scattering coefficient represents its brightness. Then this distribution of scattering coefficients can be expressed as,

\begin{equation} \mathbf{\Delta}_{P\times Q}=\begin{bmatrix}
   \mathbf{\delta}_{11} & \cdots & \mathbf{\delta}_{1Q} \\
  \vdots & \ddots & \vdots \\
   \mathbf{\delta}_{P1} & \cdots &  \mathbf{\delta}_{PQ}
\end{bmatrix}. \end{equation}

The algorithm proposed in \cite{jiying2010high} links the relationship between received signals and thermal light fields at the imaging area by transmissivity. Since both the transmissivity and scattering coefficients describe the system impulse response, the above algorithm can therefore be exploited into our proposed scenario,

\begin{gather}\label{equation1}
\begin{bmatrix}
   R_1  \\
   R_2  \\
    \vdots \\
   R_N  \\
\end{bmatrix}=\rho
\begin{bmatrix}
   E_{1,1}^{(1)},& \ldots E_{p,1}^{(1)},& \ldots E_{P,Q}^{(1)}  \\
   E_{1,1}^{(2)},& \ldots E_{p,1}^{(2)},& \ldots E_{P,Q}^{(2)}  \\
    \vdots & \vdots & \vdots\\
   E_{1,1}^{(N)},& \ldots E_{p,1}^{(N)},& \ldots E_{P,Q}^{(N)}  \\
\end{bmatrix}
\begin{bmatrix}
   \delta_{1,1}  \\
   \delta_{2,1}  \\
    \vdots \\
   \delta_{P,Q}  \\
\end{bmatrix},
\end{gather}where, $R_n(n=1,2,\ldots N, N=P\times Q)$ is the received signal of N times detection, $E_{p,q}^{(n)}$ is the EM field at one sub-plane in the $n$th detection,

\begin{equation}\label{EM}
E_{p,q}^{(n)}(t)=\sum_{i=1}^I S_{i}(t-\tau_{i}),
\end{equation}where $I$ is the total number of distributed antennas.

Equation(\ref{equation1}) can be rewritten as,

\begin{equation}
\bm{\mathfrak{R}}=\bm{\mathfrak{E}} \bm{\mathfrak{\delta}},
\end{equation}then,

\begin{equation}\label{equation3}
\bm{\mathfrak{\delta}}=\bm{\mathfrak{E}^{-1}} \bm{\mathfrak{R}}.
\end{equation}

Equation(\ref{equation3}) shows that if the matrix of EM field $\bm{\mathfrak{E}}$ at the imaging plane is full rank, there will exists a unique solution for the distribution of scattering coefficients at the target imaging plane.

Meanwhile, under low SNR level, since the noise dominates the reconstruction procedure, the reconstruction algorithm should be modified to obtain an optimal $ \bm{\mathfrak{\delta}}$ by minimizing the objective function $|| \bm{\mathfrak{R}'- \mathfrak{E} \mathfrak{\delta}} ||$, where $\bm{ \mathfrak{R}'}$ is the received signal with noise. An optimization algorithm can be used here, such as Gradient Projection (GP) algorithm\cite{figueiredo2007gradient},etc.

%==========================================================================================================================================================================
\section{Simulation Results} \label{simulation}
In the simulation, the investigation domain is at a distance of $1m$ away from the plane of distributed antennas. There are 4 transmitting antennas deployed with equal distances away from the target, forming a squared shape with a side length of $b$. The receiving antenna is set to be at the geometry center of the plane of distributed antennas. Details of detecting signals are also shown in Table. \ref{simulationSettings}.

\begin{table}[!t]
\renewcommand{\arraystretch}{1.3}
\caption{Signal Parameter Settings}
\label{simulationSettings}
\centering
\begin{tabular}{|c|c|}
\hline
\ \bfseries Bandwidth & $B=2GHz$\\
\hline
\ \bfseries Center Frequency & $f_c=1GHz$\\
\hline
\ \bfseries Pulse Width & $T_p=3\mu s$ \\
\hline
\ \bfseries Sampling Frequency & $f_s=B/2=500MHz$ \\
\hline
\ \bfseries Chirp Rate & $\gamma=B/T_p=6.67\times 10^{14}$ \\
\hline
%\ \bfseries Sampling Interval & $3ns$ \\
%\hline
%\hline
%\ \bfseries Coherence Time & $0.5ns$ \\
%\hline
%\ \bfseries Bandwidth & $2GHz$ \\
%\hline
%\ \bfseries Center Frequency & $1GHz$ \\
%\hline
\end{tabular}
\end{table}

\subsection{Simulation Results for Sampling Interval}
%When the bandwidth of the signal carrier is $2GHz$, according to equation (\ref{coherenceEqua}), the coherence time of detecting signals is equal to $0.5 ns$. The comparison shown in Figure.\ref{coherenceTimeComparison} gives a clear view that when the sampling rate of the received signal matches the Nyquist's sampling theorem, which is set to be $4GHz$ in the simulation, there exists remarkable correlation between different samplings along the time sequence. While when the sampling rate satisfies the GI required sampling rate, which is $500MHz$ in the simulation, the result shows a significant self-correlation level, indicating a full independence in time. Note that in the previous research of microwave GI implementations\cite{6518188}\cite{7031421}, the sampling rates are both set to match Nyquist's sampling theorem. However, the comparison displayed above shows that the GI sampling requirement discussed in Section \ref{theoreticalAna} can provide a better performance in ensuring the independence in time.

\begin{figure}[!htp]
\centering
    \mbox{
       \subfigure[]{\scalebox{0.2}{\includegraphics{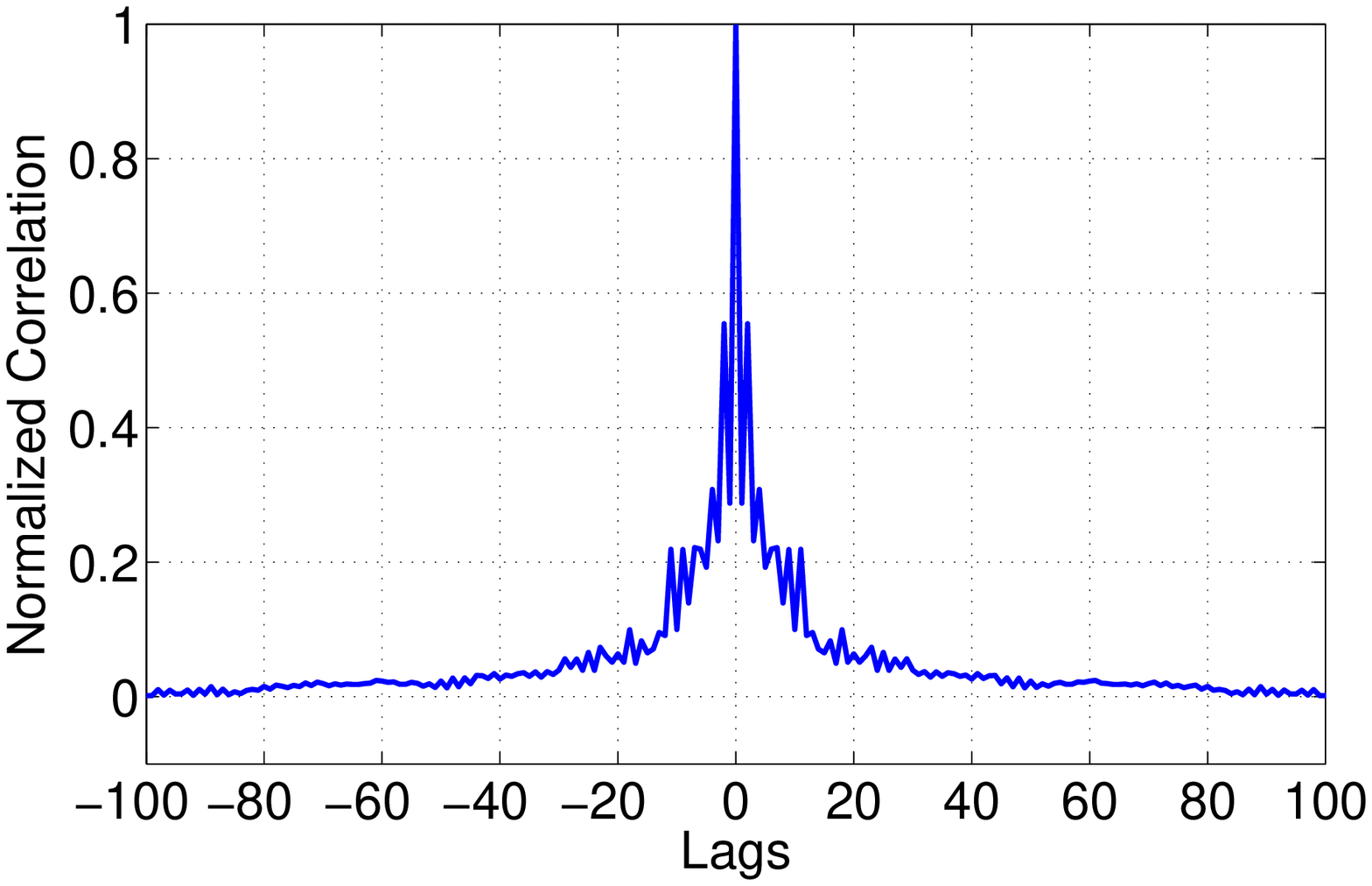}}}
       \subfigure[]{\scalebox{0.2}{\includegraphics{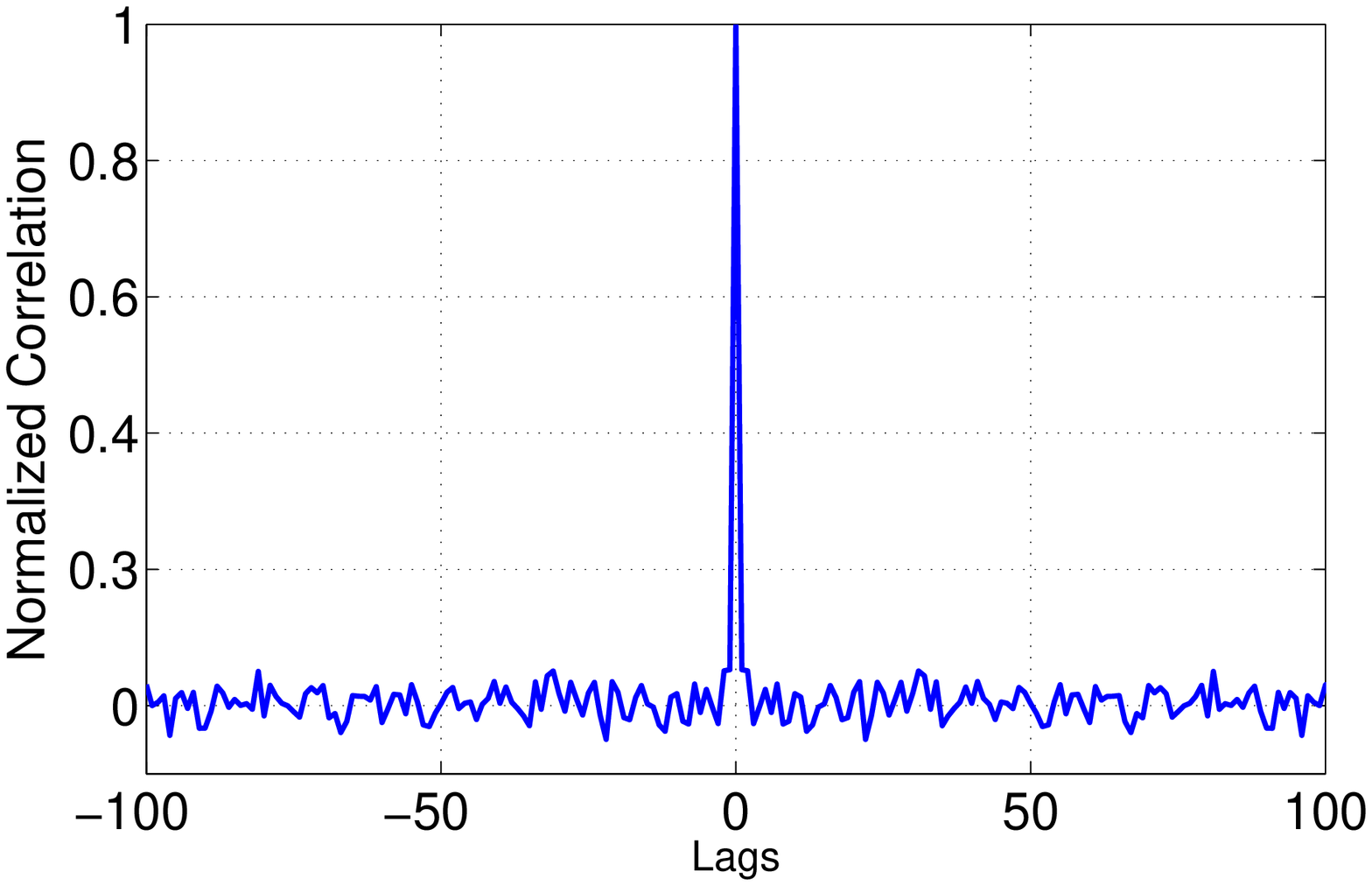}}}}
\caption{Comparison of normalized self-correlation results between Nyquist's sampling rate and GI required sampling rate. (a) Self-correlation result at Nyquist's sampling rate ($4GHz$). (b) Self-correlation result at GI required sampling rate ($500MHz$).}
\label{coherenceTimeComparison}
\end{figure}

When the bandwidth of the signal carrier is $2GHz$, according to equation (\ref{coherenceEqua}), the coherence time of detecting signals is equal to $0.5 ns$. The comparison shown in Figure.\ref{coherenceTimeComparison} gives a clear view that when the sampling rate of the received signal matches the Nyquist's sampling theorem, which is set to be $4GHz$ in the simulation, there exists remarkable correlations between different samplings along the time sequence. While when the sampling rate satisfies the GI required sampling rate, which is $500MHz$ in the simulation, the result shows a significant self-correlation level, indicating a high independence in time. This comparison shows that the GI sampling requirement discussed in Section \ref{theoreticalAna} can provide a better performance in ensuring the independence in time.

\subsection{Simulation Results for Coherence Size}
\begin{figure}[!htp]
\centering
    \mbox{
       \subfigure[]{\scalebox{0.18}{\includegraphics{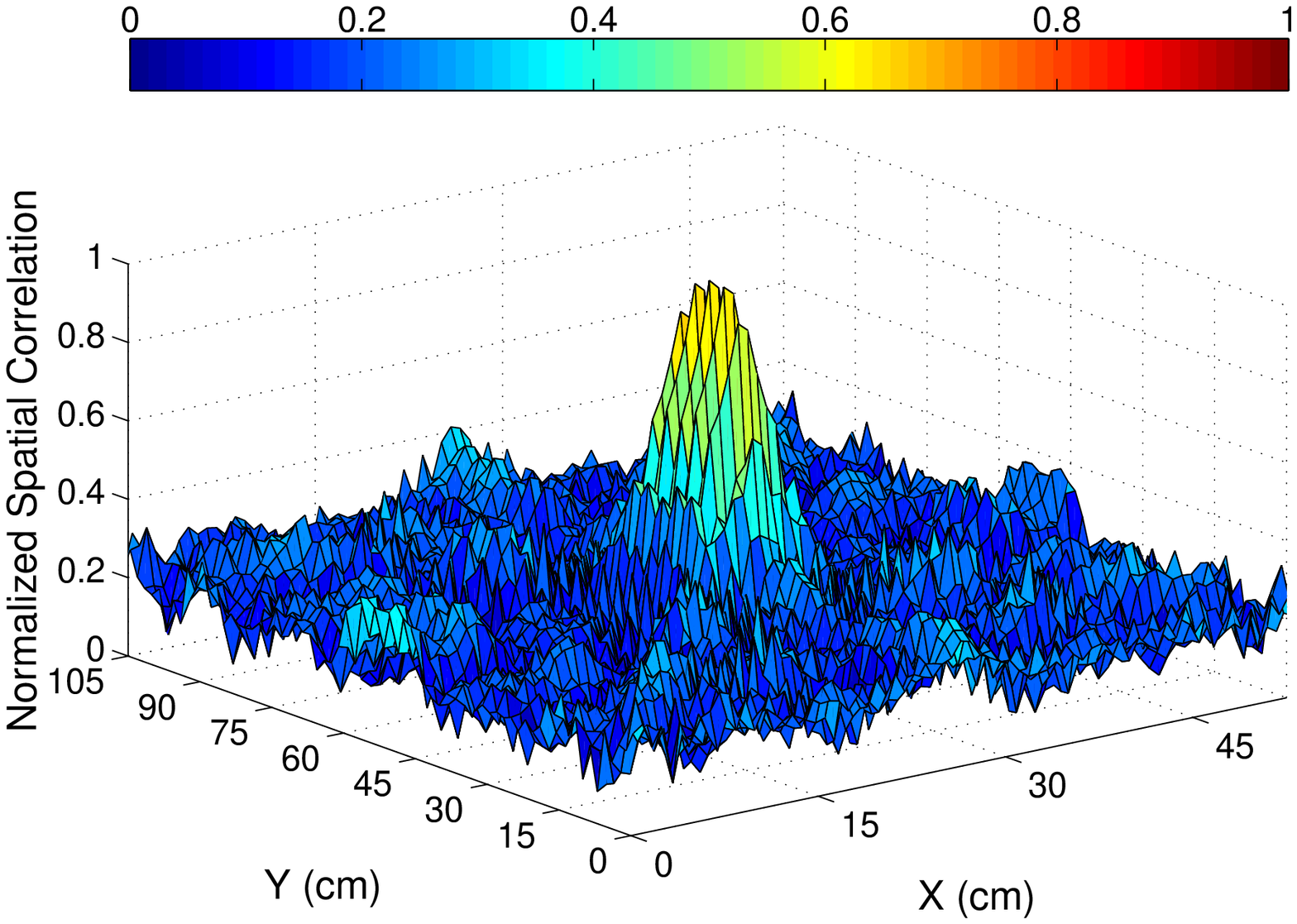}}}
       \subfigure[]{\scalebox{0.18}{\includegraphics{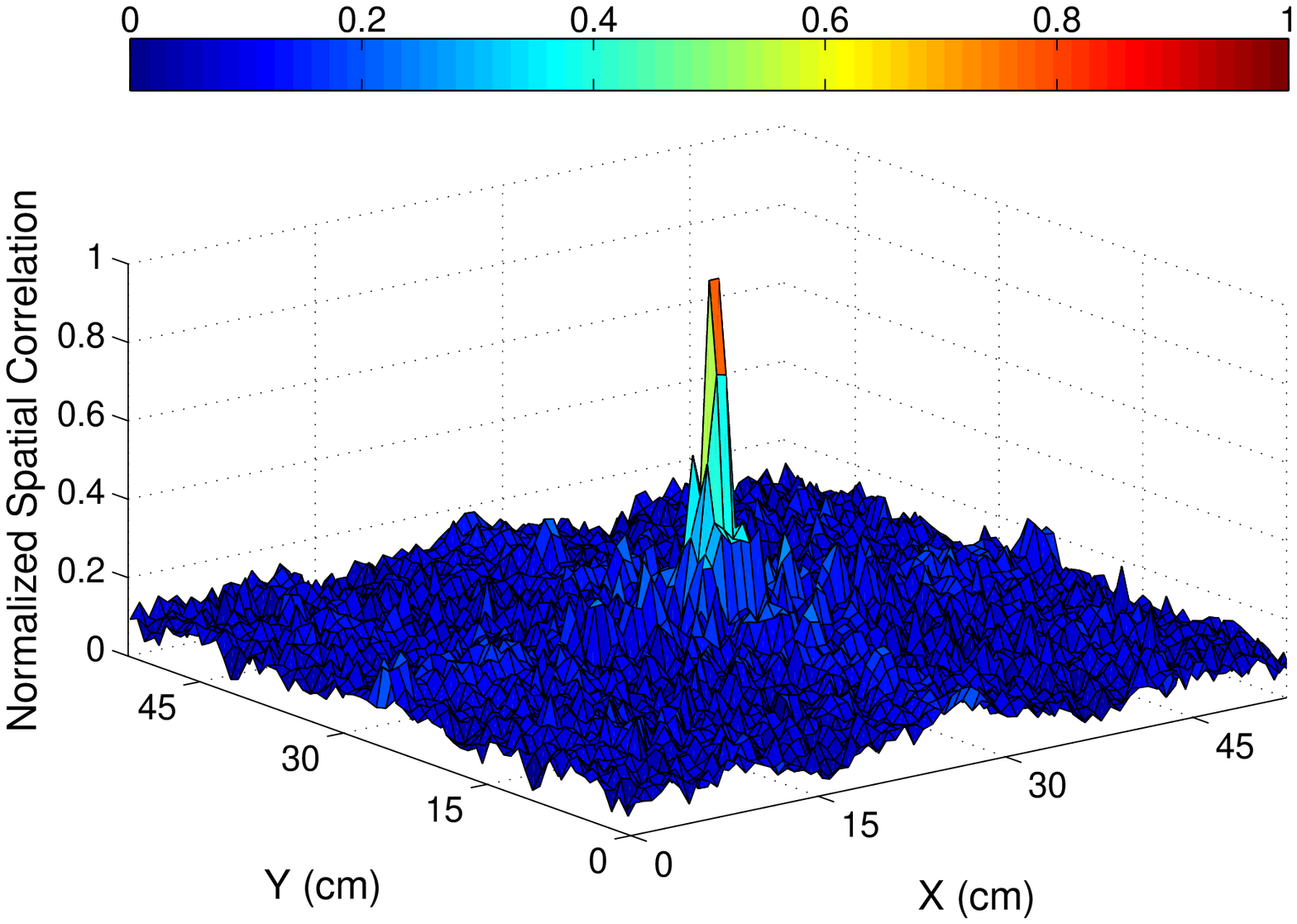}}}}
    \mbox{
       \subfigure[]{\scalebox{0.18}{\includegraphics{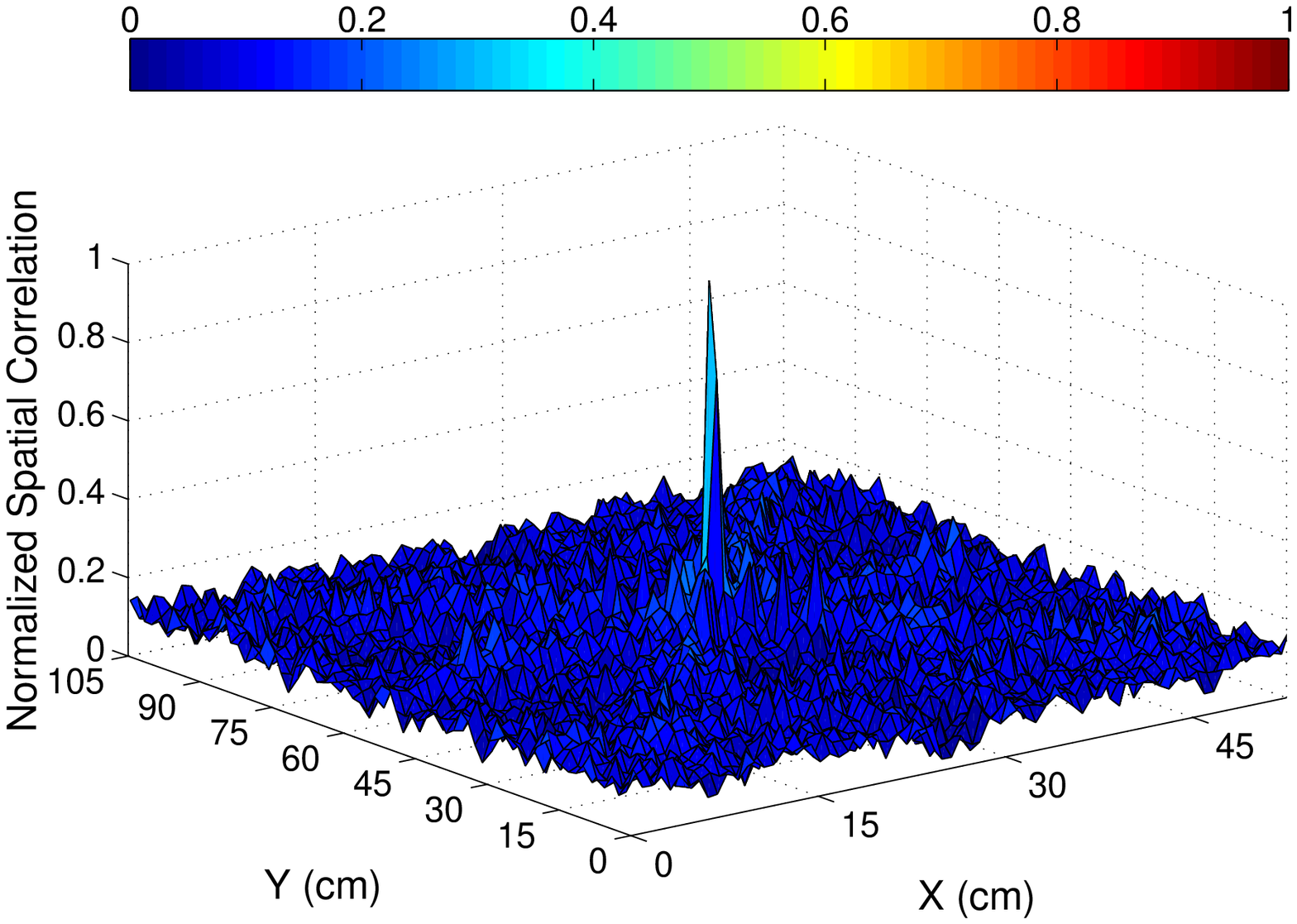}}}
       \subfigure[]{\scalebox{0.18}{\includegraphics{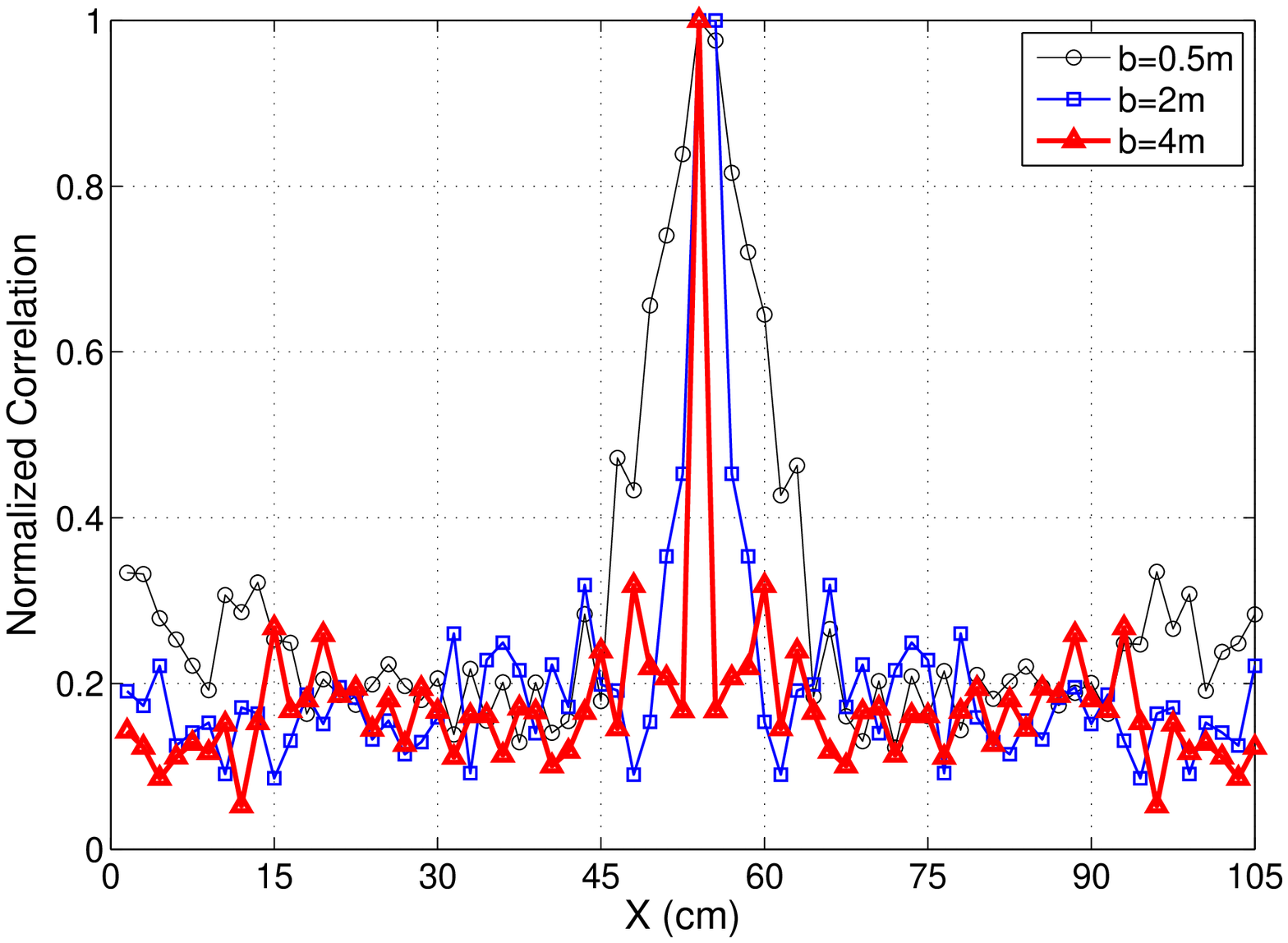}}}}
\caption{Comparison of Normalized spatial correlation between different $b$ values. (a) $b=0.5m$ (b) $b=1m$. (c) $b=4m$. (d) A comparison of cross-section profiles.}
\label{spatialCorrelation}
\end{figure}

The comparison in Figure.\ref{spatialCorrelation} shows that with the increasing of size $b$, the spatial-correlation level of EM fields between different sub-planes reduces correspondingly. This result is consistent with the conclusion which is discussed in Section \ref{theoreticalAna}.

\subsection{Simulation Results for Image Reconstruction}
\begin{figure}[!htp]
\centering
    \mbox{
       \subfigure[]{\scalebox{0.18}{\includegraphics{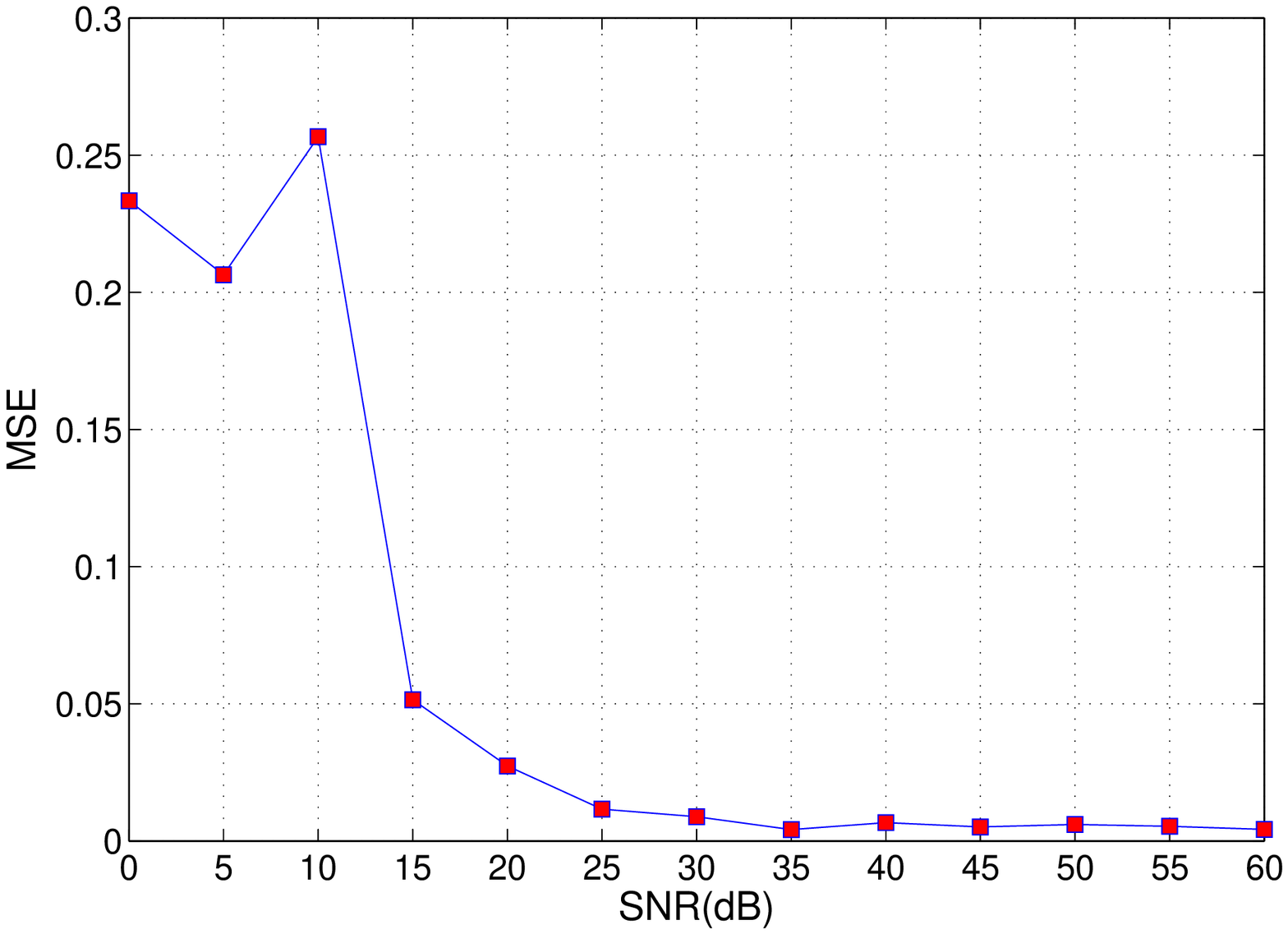}}}
       \subfigure[]{\scalebox{0.18}{\includegraphics{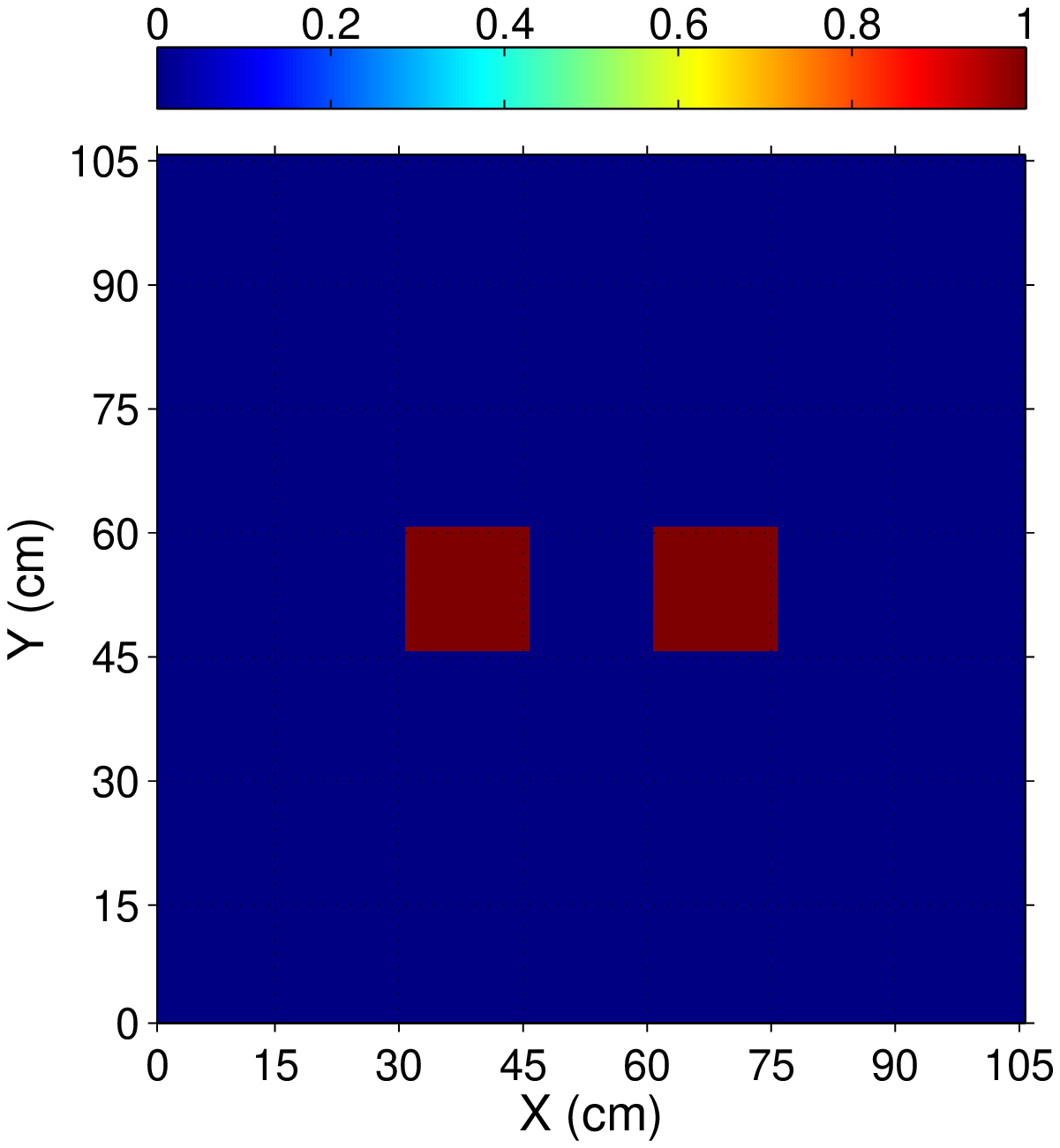}}}}
    \mbox{
       \subfigure[]{\scalebox{0.18}{\includegraphics{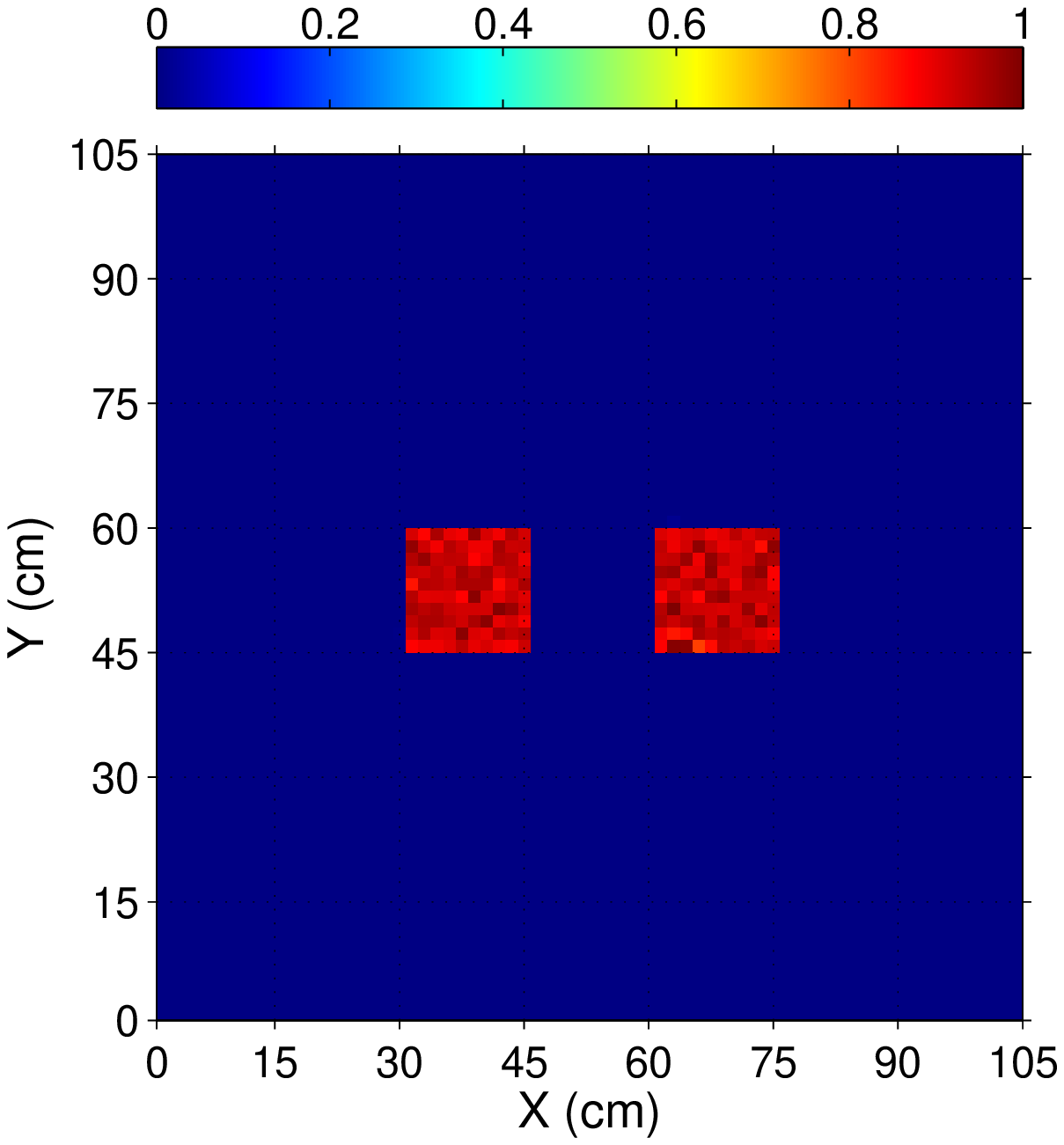}}}
       \subfigure[]{\scalebox{0.18}{\includegraphics{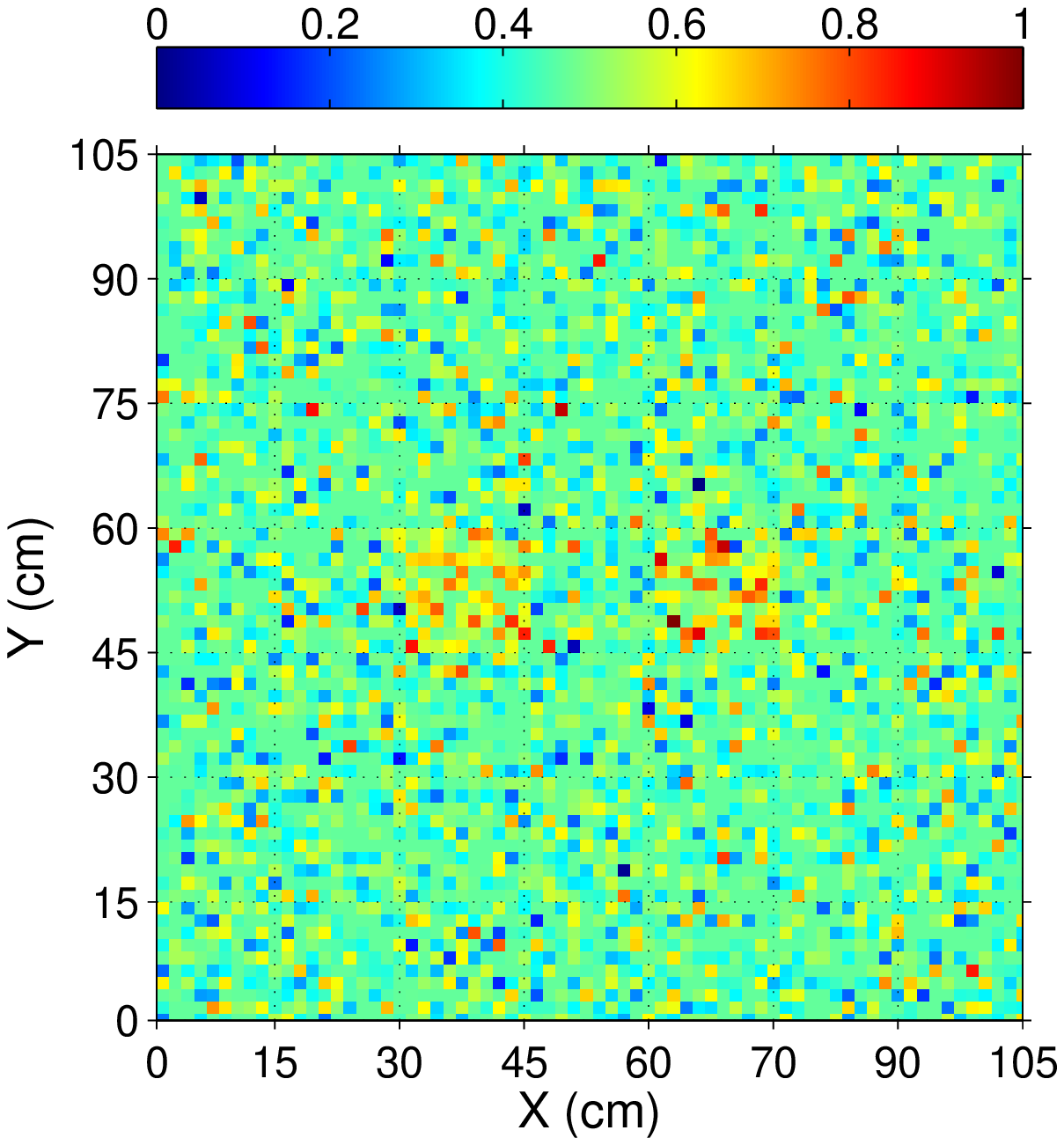}}}}
\caption{Reconstruction results with $b=4m$ and a discretization size of $15mm\times15mm$. (a) Reconstruction MSE under different SNR levels. (b) The original scenario. (c) Reconstructed result under SNR=30. (d) Reconstructed result under SNR=0.}
\label{reconstructionComparison}
\end{figure}

As shown in Figure.\ref{reconstructionComparison} (b), two identical targets with the same size, shape and contrast level are deployed in the scenario for further investigations about the reconstruction performance. Mean Square Error (MSE) is introduced here to qualitatively evaluate the reconstruction error. Figure.\ref{reconstructionComparison} (a) shows the MSE results under different SNR levels, where we can conclude that with the increasing of SNR, the reconstruction MSE will decrease dramatically at first and remain stable at high SNR levels. Furthermore, as we can see from Figure.\ref{reconstructionComparison} (c) and (d), which demonstrate the reconstructed results under SNR=30dB and SNR=0dB respectively, our proposed GI and distributed antennas based microwave surveillance method can effectively retrieve the target image, and ensure a high quality result with clear edges under high SNR conditions.

%In the simulation of image reconstruction using the proposed scheme and settings of parameters, we applied the EM fields collected by XFDTD at the target plane as detecting fields, and a grey-scale picture as the distribution of target reflection coefficients to simulate the detecting procedure. The original image and the distribution of the reflectivity represented by the grey-scale picture is shown in Figure.\ref{reconstructionComparison} (a) and (b), and the corresponding reconstruction results are displayed in Figure.\ref{reconstructionComparison} (c) and (d). The comparison of the simulation results indicates that the proposed microwave GI and distributed antennas based massive surveillance scheme can effectively retrieve the target image with high quality. 

%==================================================================================================================
\section{Conclusion} \label{conclusion}
In this paper, we proposed a novel surveillance method based on microwave GI and distributed antennas. To our best knowledge, this is the first time for GI to be introduced into microwave surveillance applications. It is also the first approach in presenting clear theoretical analysis in the sampling requirement of microwave GI and the influence factors in its imaging resolution. Validation via simulations shows that the proposed method can provide high-quality surveillance performance with low system complexity.

\bibliographystyle{IEEEbib}
\bibliography{Journal_main}
\end{document}